# Critique of Optical Negative Refraction Superlensing

George Christou and Christos Mias[1§]

[1]School of Engineering, Warwick University, Coventry, CV4 7AL, UK

**Abstract**

Has the ten-year old quest for the optical superlens, based on Veselago's hypothesis of negative refraction, been a chimera? We argue that Pendry's alternative prescription of the silver superlens is nothing more than an application of the natural phenomenon of surface plasmon resonance that occurs in the noble metal films. This phenomenon does not predict the reality of Veselago's negative refractive index materials. We give a simple explanation of how this resonance achieves a field intensity enhancement at the interface of silver and air without involving the concept of negative refraction.

---

[§] Corresponding author: christos.mias@warwick.ac.uk

Pendry [1] states: "a rigorous solution of Maxwell's equations leads to the remarkable conclusion that the near fields can be focused by a negative refraction index material as long as $\varepsilon = -1$ and $\mu = -1$". This material with $\varepsilon = \mu = -1$ is his conceptual superlens. He also points out that this is the condition for surface plasmons to exist at optical frequencies. These surface plasmons on the front surface of a slab of this material are excited by a very weak evanescent near field of an object. These excited surface plasmons begin a resonant electrostatic interaction that excites surface plasmons on the rear surface of the slab. Since this resonance is very strong, the image is transferred to the rear surface in a hugely amplified form [1]. He presents the above schematically in ray form in figure 4c and with an intensity profile in figure 4d of reference [2].

Unable to realise Veselago's hypothetical negative refractive materials [3] with $\varepsilon = \mu = -1$ for his superlens, Pendry, disregarding $\mu$ at the optical frequencies, turned to the natural material silver [4] with its well known optical characteristics and proposed the silver superlens which he considers its approximate equivalent at optical frequencies. He stated that the best value he could get for the silver relative permittivity at these frequencies is $\varepsilon = -1 + 0.4i$. In [5] he claims that two groups [6]-[7] practically demonstrated negative refraction superlensing in an optical system. These groups used a layer of silver of about 40 nm thick to image 365 nm wavelength light emanating from shaped apertures smaller than the light's wavelength. We suggest that what they really demonstrated is the phenomenon of surface plasmon resonance (SPR) in silver. This phenomenon takes place when the wave vector of an incident evanescent wave matches the wave vector of the surface plasmons. When this occurs, there is a resonant power transfer and a field intensity enhancement. Before considering further the SPR phenomenon it is proper to mention that it has been known [8] and exploited [9] for decades. Among the many worth noting industrial applications, we mention its successful exploitation in some optical sensors used in the biological and chemical fields. The detection of molecules is done through their interaction with the resonantly enhanced evanescent field at the metal-dielectric interface [9]. Needless to stress that for this evanescent field enhancement an incident evanescent field is a prerequisite.

Pendry suggests that to focus the evanescent near fields we need to amplify them to compensate for their rapid decay. He depicts schematically in figure 6 [1] this focusing by first amplifying them exponentially through the resonant interaction of the surface plasmons on both sides of a slab ($n = -1$). To the best of our knowledge he does not present anywhere the workings of his suggested silver superlens. But being realists, we are interested in what happens in the silver film. We provide a simple explanation of why there is an enhanced field at the silver film – air interface. We realised that what happens during the SPR interaction is the energy transfer between the incident photons of large wavelength, less than 1 μm, and the silver free conduction electrons of small wavelength, less than 1 nm. Needless to say that the generated surface plasmons and the incident photons have the same frequency. The resulting energy concentration in the surface plasmons at resonance produces an enhancement of the evanescent field intensity at the rear interface. To emphasise the importance of resonance for this field intensity enhancement we show in fig. 1 the spatial shape of the field intensity across the silver film optimum thickness for two cases: at resonance and off resonance. More on surface plasmons later.

At this point we must explain what is meant by field intensity enhancement. However, before doing so it is interesting to be reminded of other electromagnetic resonances, say the simple a.c. series resonance. In the series resonance we have the circuit magnification factor Q which is the ratio of the much greater $V_L$ (or $V_C$) voltage to the supply voltage. For the SPR case let us consider the energy of the large-wavelength incident photons of an evanescent field, on one unit area of the silver film surface. Because of the photons energy we have a certain field intensity. During the SPR interaction this energy is transferred to free conduction electrons of smaller wavelength. Thus surface plasmons are created on the surfaces of the film. The creation of surface plasmons entails the compression of the electromagnetic wave of the incident photons into a longitudinal surface plasmon wave. The result is a much greater field intensity on a unit area on the air side of the silver film. The field intensity enhancement factor is the ratio of the resulting field intensity at the interface to the incident field intensity. For the resonant case shown in fig. 1, an enhancement factor of 48 is achieved. Summing up this simple treatment of the field intensity enhancement we stress again that it is the result of the energy transfer at SPR between the incident photons and the free conduction electrons. We must also emphasise that the SPR phenomenon only permits an energy transfer from photons to plasmons when both the momentum and energy of the incident photons match exactly the momentum and energy of the plasmons. We repeat again that the Zhang group [7] exploited this SPR phenomenon to transfer energy from an incident evanescent field through a silver film and recorded it on a layer of photoresist. The incident evanescent field resulted from the passage of 365 nm wavelength light through apertures, smaller than the light's wavelength, in a 50 nm thick chromium film.

Now let us see what happens at the air side of the silver film at SPR and how the enhanced evanescent field is set up. We need to elaborate a bit more on the nature of surface plasmons. We know that they are optically induced oscillations of the free conduction electrons at the surface of a metal. They can exist at the interface of materials such as silver and air whose dielectric constants are of opposite signs. The surface plasmons are longitudinal waves that propagate along the silver surface. They are eventually absorbed. Their propagation length is about 20 µm at the wavelength of 500 nm. However, because of their movement, they set up in air a vertical evanescent field which, at resonance, is enhanced at the silver film - air interface but its amplitude exponentially decays within a vertical distance of about 250 nm, half the wavelength of the light involved [10]. We emphasise that this bounded evanescent field, although decaying exponentially, remains vertical; it does not bend. There is no indication that the silver film refracts negatively. From the description of the superlens in the box on page 48 [5] the silver film cannot be considered as an optical superlens. We must stress again that the realised field intensity enhancement at SPR is due to the nature of the surface plasmons and the surface plasmons are neither the result of negative refraction nor do they produce the cherished negative refraction. There is no need to rename the SPR phenomenon as negative refraction superlensing.

In conclusion, one can neither consider the occurrence of a field intensity enhancement at the silver film - air interface as a prediction for the reality of Veselago's negative refractive index materials nor can one agree with Pendry's assertion that the prospect of negative refraction has caused physicists to re-examine virtually all of electromagnetics [5]. However, "everything changes" (Heraclitus 536 – 470 B.C.). Only time, through real applications, will be the ultimate judge for the acceptance of the optical negative refraction superlens hypothesis in the theory of electromagnetics. In the mean-time we hope more researchers will become interested in the promising field of plasmonics. Plasmonics aspires to integrate the micron-scale photonics to the nano-scale electronics. Epigrammatically we express our hope that in ten years time we shall witness the birth of a plethora of plasmonic components from the marriage of electronics and photonics.

**List of Captions**

**Figure 1**

Caption: Illustration of the spatial shape of the field intensity across the silver film optimum thickness of 60 nm for in- and off-resonance. Data obtained from fig. 2.12 of [8].

**Figure 1**

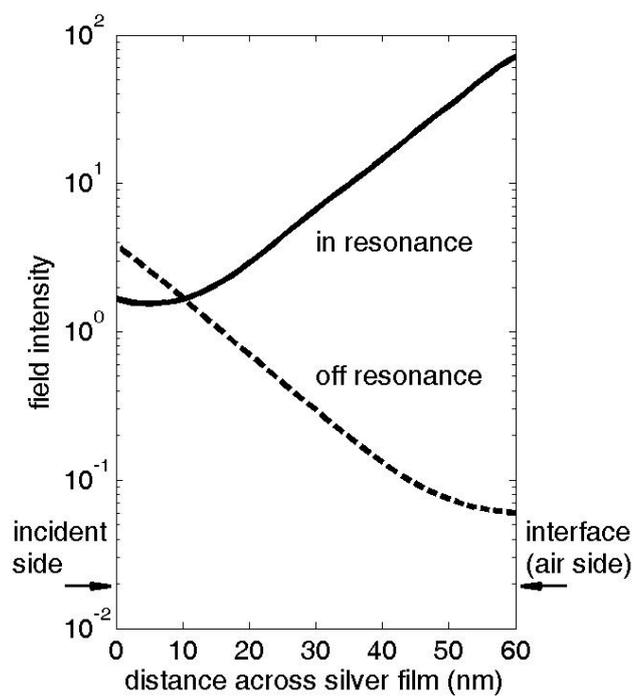